\documentclass{aa}
\usepackage{amsmath}
\usepackage{booktabs}
\usepackage{comment}
\usepackage[normalem]{ulem}

\usepackage{placeins} 

\usepackage{longtable}
\usepackage{supertabular} 
\usepackage{xcolor}

\usepackage{array}
\usepackage{natbib}
\bibpunct{(}{)}{;}{a}{}{,}
\usepackage{hypernat} % Ensures compatibility between natbib and hyperref
\usepackage[colorlinks,linkcolor=blue,citecolor=blue,urlcolor=blue]{hyperref}

\begin{document} 
%TC:ignore

\title{The 3D morphology of open clusters in the solar neighborhood III: Fractal dimension}

\subtitle{Fractal dimension of open clusters}

\author{Chang Qin \inst{1,2} \orcid{0009-0000-5426-9409}
\and Xiaoying Pang \inst{1,3}  \fnmsep\thanks{\email{Xiaoying.Pang@xjtlu.edu.cn}} \orcid{0000-0003-3389-2263}
\and Mario Pasquato \inst{4} \orcid{0000-0003-3784-5245}
\and M.B.N. Kouwenhoven \inst{1} \orcid{0000-0002-1805-0570}
\and \\
Antonella Vallenari \inst{5} \orcid{0000-0003-0014-519X}
}

\institute{
Department of Physics, Xi'an Jiaotong-Liverpool University, 111 Ren’ai Road, Dushu Lake Science and Education Innovation District, Suzhou 215123, Jiangsu Province, P.R. China. 
%\email{Xiaoying.Pang@xjtlu.edu.cn}
\and 
Key Laboratory of Quark and Lepton Physics (MOE) and Institute of Particle Physics, Central China Normal University, Wuhan 430079, P.R. China.
\and
Shanghai Key Laboratory for Astrophysics, Shanghai Normal University, 100 Guilin Road, Shanghai 200234, P.R. China
\and
IASF Milano, via Alfonso Corti 12, Milano, Italy
\and 
INAF-Osservatorio Astronomico di Padova, Vicolo Osservatorio 5, 35122 Padova, Italy
}

\titlerunning{Fractal dimension of open clusters}
\authorrunning{Qin et al.}

\date{Received Octobor 3, 2024; Accepted December 14, 2024}

%-------------------------------------------------------------------------------------------% 
%*******************************************************************************************%

\abstract{
We analyze the fractal dimension of open clusters using 3D spatial data from Gaia DR\,3 for 93 open clusters from \citet{Pang_2024} and 127 open clusters from \citet[][]{Hunt2024} within 500~pc. The box-counting method is adopted to calculate the fractal dimension of each cluster in three regions: the all-member region, $r \leq r_t$ (inside the tidal radius), and $r > r_t$ (outside the tidal radius). In both the Pang and Hunt catalogs, the fractal dimensions are smaller for the regions $r > r_t$ than those for $r \leq r_t$, indicating that the stellar distribution is more clumpy in the cluster outskirts. We classify cluster morphology based on the fractal dimension via the Gaussian Mixture Model. Our study shows that the fractal dimension can efficiently classify clusters in the Pang catalog into two groups. The fractal dimension of the clusters in the Pang catalog declines with age, which is attributed to the development of tidal tails. This is consistent with the expectations from the dynamical evolution of open clusters. We find strong evidence that the fractal dimension increases with cluster mass, which implies that higher-mass clusters are formed hierarchically from the mergers of lower-mass filamentary-type stellar groups. The transition of the fractal dimension for the spatial distribution of open clusters provides a useful tool to trace the Galactic star forming structures, from the location of the Local Bubble within the solar neighborhood to the spiral arms across the Galaxy.}

%TC:endignore

\maketitle

%-------------------------------------------------------------------------------------------%
%*******************************************************************************************%
\section{Introduction}\label{sec:intro}

Open clusters are formed from interstellar gas in dense molecular clouds \citep[e.g.,][]{lada2003}. The spatial distribution of member stars in open clusters changes over time as clusters evolve. Quantifying the spatial distribution can thus provide valuable insights into the formation process and the subsequent dynamical evolution of open clusters. 

% young open clusters:
According to the theory of hierarchical star formation of \citet{kruijssen2012}, gravitationally-bound young open clusters (with ages $\leq 100$~Myr) are primarily formed in high-density regions \citep{vazquez-semadeni_hierarchical_2016, trevino-morales2019, ward_not_2020}. Such clusters exhibit fractal substructures, as predicted by the conveyor belt mechanism \citep{clarke2010, arnold2017, fujii2021}. On the contrary, clusters formed in low-density (filamentary) regions are often characterized by filamentary substructures, and such clusters disperse rapidly after gas removal. 

% old open clusters:
Internal two-body relaxation and the influence of the external Galactic tides play an important role in the evolution of older open clusters (with ages exceeding $100$~Myr). Internal two-body relaxation shapes open clusters with a dense core, accompanied by a lower-density halo \citep{pang2022a}. When Galactic tidal forces are substantial, elongated tidal-tail substructures tend to emerge over time \citep{Roeser2019, tang2019,pang2021a, pang2022a}.

% how morphology was studied in the past:
A natural first approach to studying the morphology of open clusters is through visual inspection, although this method lacks accuracy and reproducibility. Another popular method to quantify 2D morphology is calculating the radial density profile of star clusters and then fitting the profile to various models, such as the Elson-Fall-Freeman (EFF) model \citep{elson1987} or the King model \citep{1962AJ.....67..471K}. The aforementioned pioneering works on the morphology of star clusters only used the 2D (projected) distribution of member stars.

% how gaia can help morphology determination:
With the release of Gaia data \citep{2016A&A...595A...1G}, morphology research shifted from 2D spacial data to 3D spatial data, which is attributed to the significantly improved parallax measurement. \citet{pang2021a} introduced a quantitative approach to determine the 3D morphology for the region within the tidal radius of the cluster, using the ellipsoid fitting, which is an advanced version for estimating the ellipticity of a 2D projected distribution \citep{chen2004, tarricq2022}. However, this method is sensitive to extended outliers and is therefore mostly useful for the dense, bound regions of open clusters. To better distinguish between the extended substructures outside tidal radius in star clusters into different types, visual inspection was used again in \citet{pang2022a} to carry out the classification. \citet{pang2022a} successfully classified open clusters into filamentary, fractal, halo, and tidal-tail types. However, only 53 out of 85 open clusters in their catalog have been classified. The complexity of the 3D structure of a star cluster cannot be easily analyzed by the human eye alone. A quantitative approach is required to objectively and accurately determine the 3D morphology of star clusters.

% description of what is new in this paper:
In this paper, we quantify the morphology of open clusters in the solar neighborhood by introducing a quantity that does not depend on the density profile. The fractal dimension \citep{10.1119/1.13295} is measured for characterizing shapes by quantifying the complexity. The fractal dimension has been widely applied in scientific research. 
In astronomy, for example, the fractal dimension can be used to study the large-scale distribution of galaxies \citep{ribeiro_miguelote_1998, Elmegreen2001}. \citet{1987A&A...179..249F} calculated the fractal dimension of star-forming sites in 19 spiral galaxies for classification. 
The concept of fractal dimension has also been used to study turbulence in Giant HII Regions \citep{Caicedo-Ortiz_2015}, and the distribution of the interstellar medium \citep{Sánchez_2005}. The fractal dimension was first used to examine the spatial distribution of young open cluster samples in the solar neighborhood by \citet{2006A&A...452..163D}. In the latter study, each open cluster was considered as a point. Our study is the first to calculate the fractal dimension for individual open clusters in the solar neighborhood based on the 3D positions of member stars obtained from Gaia DR\,3 data using the box-counting method \citep{GRASSBERGER1983224}. 

% structure of the paper:
This paper is organized as follows. In Sect.~\ref{sec:gaia}, we introduce the two open cluster catalogs. The fractal dimension is described in Sect.~\ref{sec:fdim}, with Sect.~\ref{sec:fdim_comp} detailing its computation and Sect.~\ref{sec:fdim_uncertainty} addressing its uncertainty. In Sect.~\ref{sec:fdim_solar}, we analyze the fractal dimension in different regions of each cluster (Sect.~\ref{sec:fdim_dist_solar}) and discuss the fractal dimensions of different morphological types in Sect.~\ref{sec:fdim_type_solar}. A classification based on fractal dimension is performed in Sect.~\ref{sec:fdim_classify}. Sect.~\ref{sec:fdim_discuss} provides discussions on the fractal dimension in relation to open cluster dynamical evolution (Sect.~\ref{sec:fdim_dynamical}), star formation within open clusters (Sect.~\ref{sec:fdim_starformation}), and Galactic structures (Sect.~\ref{sec:fdim_galac}). Finally, we provide a brief summary of our findings in Sect.~\ref{sec:sum}.

%%%%%%%%%%%%%%%%%%%%%%%%%%%%%%%%%%%%%%%%
%%%%%%%%%%%%%%%%%%%%%%%%%%%%%%%%%%%%%%%%
%%%%%%%%%%%%%%%%%%%%%%%%%%%%%%%%%%%%%%%%

\section{Open clusters catalogs}\label{sec:gaia}

% introduction paragraph:
Gaia DR\,3 \citep{gaia_collaboration_gaia_2022} has offered unprecedented high-accuracy positions on the sky and parallax for more than $1.8$ billion sources, and so has Gaia EDR\,3 \citep{gaia2021}. In this study, we use the catalog based on Gaia DR\,3 \citep{gaia_collaboration_gaia_2022}: the Pang star cluster catalog  \citep{Pang_2024}, and compare the findings with those of the Hunt star cluster catalog \citep{Hunt2024}.

% introduce pang catalog data:
We investigate star clusters within 500~pc in the solar neighborhood from the catalog of \citet{Pang_2024}, based on $(X,Y,Z)$ coordinates of the member stars of 93~star clusters. \citet{pang2021a, pang2021b, li2021, pang2022a, pang2022c, Pang_2024} identified member stars of a total 93~star clusters. Their study strictly requires member stars to have measurement uncertainties below $10\%$ in their parallax and photometry. The machine learning algorithm Stars’ Galactic Origin \citep[StarGO]{yuan2018} is used to select star cluster members based on Gaia EDR\,3 and DR\,3 data \citep{gaia2021,gaia_collaboration_gaia_2022}. The field star contamination rate was kept within a $5\%$ threshold, corresponding to a membership probability of $95\%$. 
The heliocentric Cartesian coordinates of member stars in these open clusters are adjusted based on the distance correction using the Bayesian approach proposed in \citet{carrera2019}, which assumes a two-component prior. A normal distribution prior is assumed for the individual distances to the cluster stars, while an exponentially decreasing profile prior is used for the distances to field stars \cite{bailer2015}. The corrected distance for each star is derived from the mean of the posterior distribution. No distance correction is applied to cluster Group~X, as it lacks a central concentration of stars and instead exhibits a two-piece fragmented spatial distribution. Group X is currently in the final stages of disruption \citep{tang2019}.

% introduce hunt catalog data:
 For comparison with the Pang catalog clusters, we extend our investigation to open clusters from the all-sky cluster catalog of \citet{Hunt2024} with distances below 500~pc. We further select open clusters with a median color-magnitude diagram (CMD) class above $0.5$ and a Cluster Significance Test (CST) score greater than $5\sigma$ \citep[see][for details]{Hunt2023}. As for the Pang catalog clusters, only member stars with parallaxes and photometric measurements within a $10\%$ uncertainty \citep[][Appendix~C]{lindegren2018} are included for further analysis. We obtain a total of 127 open clusters. The distances of individual member stars are corrected using the same Bayesian approach \citep{carrera2019} that is applied to the catalog of \cite{Pang_2024}.

% summary paragraph about data:
In total, the fractal dimension analysis includes 93~open clusters from \citet{Pang_2024} (hereafter: the Pang catalog clusters) and 127~open clusters from \citet{Hunt2024} (hereafter: the Hunt catalog clusters) mainly within 500~pc in the solar neighborhood. The basic properties for these two samples of clusters are listed in Tables~\ref{tab:fdim_Pang} and~\ref{tab:fdim_Hunt_10}.

%%%%%%%%%%%%%%%%%%%%%%%%%%%%%%%%%%%%%%%%
%%%%%%%%%%%%%%%%%%%%%%%%%%%%%%%%%%%%%%%%
%%%%%%%%%%%%%%%%%%%%%%%%%%%%%%%%%%%%%%%%

\section{Fractal dimension}\label{sec:fdim}

\subsection{Computation of fractal dimension} \label{sec:fdim_comp}

% introduce fractal dimension and usefulness:
We quantify the morphological complexity of open clusters using the fractal dimension, which serves as a measure of fractal structures. The fractal dimension offers a quantitative tool to study the spatial distribution of individual stars inside the cluster. A lower value of fractal dimension indicates a more clumpy (substructured) morphology, and vice versa.

% introduce our method for computing fdim:
In this work, we consider the box-counting dimension, also known as the Minkowski-Bouligand dimension \citep{10.5555/994632}, which can be obtained using the box-counting method \citep{GRASSBERGER1983224}. The box-counting method covers the dataset with boxes of varying sizes and counts the boxes needed to cover the entire dataset. The 3D coordinates of the member stars are standardized so that they have zero mean and unit variance before the start of box-counting, which guarantees the coordinates become all dimensionless as well. Thus, the box length $L$ below is also dimensionless. 

\begin{figure}[tb!]
\centering
\resizebox{0.85\hsize}{!}{\includegraphics{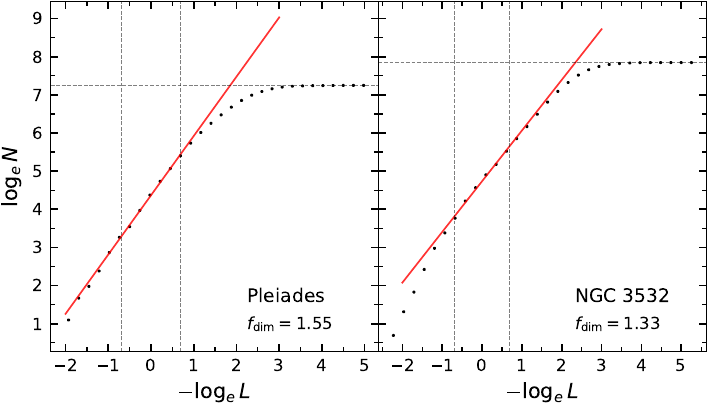}}
    \caption{The box-counting $\log_{e} N$  vs. $-\log_{e} L$ of Pleiades and NGC 3532. The slopes of the red solid lines fitted by robust regression represent the estimated fractal dimension $f_{\rm dim}$. The intervals defined by the vertical dashed lines correspond to the fitting range $[-\log_{e} 2, \log_{e} 2]$. The horizontal dashed line in each panel indicates the natural logarithm of the total number of member stars in the respective star clusters.}
\label{fig:fdim_examples}
\end{figure}

The fractal dimension $f_{\rm dim}$ via the box-counting method is defined as 
\begin{equation}\label{eq:box-counting}
    f_{\rm dim} = 
    -\frac{d \log_{e} N(L)}{d \log_{e} L},
\end{equation}
where $L$ is the length of the box and $N(L)$ is the number of boxes counted to cover the spatial distribution of member stars corresponding to the given $L$. Given a finite dataset of discrete points, this derivative must clearly be estimated using an approximate prescription, such as taking a finite difference. Moreover, its value may change as a function of box size. In our case we estimate $f_{\rm dim}$ as the slope that is obtained by linear regression between $\log_{e} N(L)$ and $-\log_{e} L$ (see Fig.~\ref{fig:fdim_examples}).

% describe the figure:
Fig.~\ref{fig:fdim_examples} shows that plateaus emerge when the number of boxes $N$ approaches the number of the member stars (horizontal dashed lines in Fig.~\ref{fig:fdim_examples}). To mitigate the bias on the plateaus and automate the computation process, we define a fixed fitting range of $[-\log_{e} 2, \log_{e} 2]$ (indicated by the vertical dashed lines in Fig.~\ref{fig:fdim_examples}). This is a range in scales equal to $1/2$ and $2$ times the scale set by the standard deviation of the coordinates, which is roughly the half-mass radius.

\subsection{Uncertainty estimation} \label{sec:fdim_uncertainty}

Here we discuss the dependence of the uncertainty in the fractal dimension on (1) observational errors in the Gaia data, (2) the number of member stars in each cluster $n_s$, and (3) the choice of contamination rate.

Gaia's astrometric uncertainties (R.A., Decl., and parallax) cause uncertainties in the 3D positions of individual stars, which can affect the derived fractal dimensions. To mitigate this, we re-assign the position of each member star in the Pang catalog, by drawing from a Gaussian distribution centered at the observed position, with the dispersion based on the observational uncertainty. The fractal dimension of the cluster is then recalculated with the newly assigned stellar positions. This process is repeated 1000 times for each cluster. The relative error of fractal dimension associated with astrometric uncertainty is only a few percent (2--6\%), which is considered acceptable. 

\begin{figure*}[tb!]
\centering
\resizebox{0.85\hsize}{!}{\includegraphics{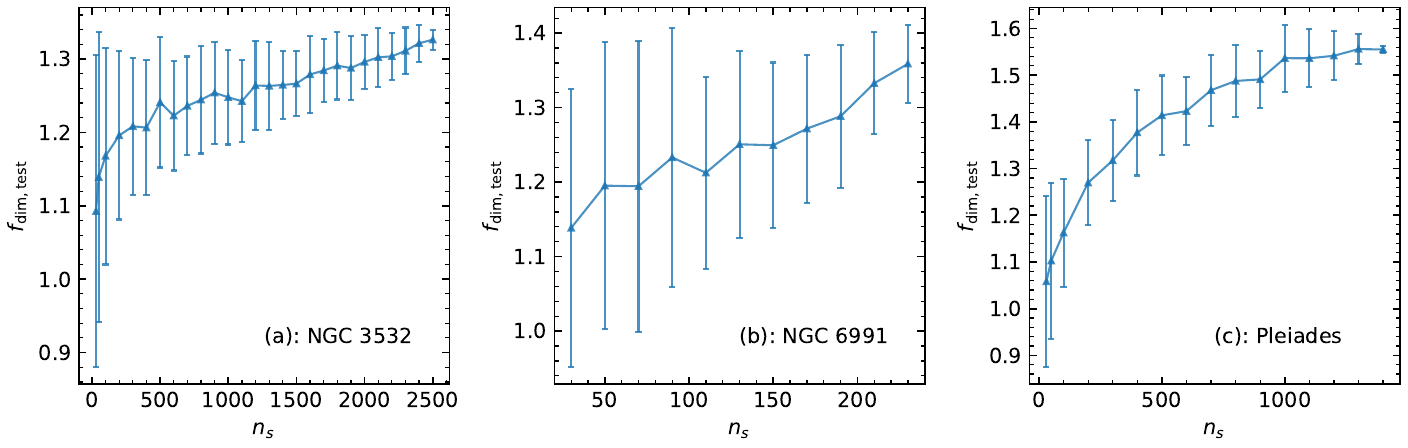}}
    \caption{Dependence of the fractal dimension $f_{\rm dim, test}$ on the number of member stars, $n_s$, for three example clusters: NGC 3532 (a), NGC 6991 (b), and  Pleiades (c). The blue triangles and the error bars indicate the corresponding mean fractal dimension and the standard deviation in the corresponding $n_s$ group of 100 iterations. }
\label{fig:fdim_sampled_num}
\end{figure*}

To estimate the effect of the member number, $n_s$, on the fractal dimension, we select three example clusters with total member numbers ranging from 240 to above 2500, NGC 3532, NGC 6991, and the Pleiades, and randomly resample different numbers of member stars (without replacement) to recalculate the value of $f_{\rm dim}$. 
For different values of $n_s \in [30, 50, 100, 200, \ldots]$ (except for NGC 6991, which has approximately 240 stars, for which an interval of 20 is used), 100 iterations are performed. The corresponding mean fractal dimension and standard deviation of these 100 trails are computed and presented in Fig.~\ref{fig:fdim_sampled_num}~(a) to (c). The fractal dimension of each of these three clusters increases with growing member number $n_s$, while the relative error (standard deviation, blue error bars) decreases with $n_s$. The relative error varies from $19\%$ (NGC 3532 with $n_s = 30$) to $0.4\%$ (Pleiades with $n_s = 1400$).

\begin{figure}[tb!]
\centering
\resizebox{0.6\hsize}{!}{\includegraphics{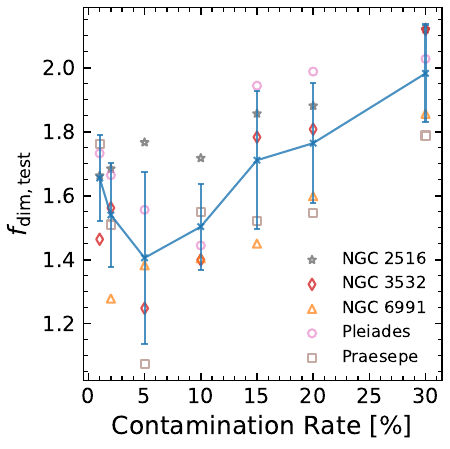}}
    \caption{Fractal dimension vs. contamination rate for five example clusters, NGC 2516 (gray stars), NGC 3532 (red diamonds), NGC 6991 (orange triangles), Pleiades (pink circles), and Praesepe (brown squares). The blue crosses and the error bars are the mean fractal dimension and the standard deviation of five clusters at the same contamination rate. }
\label{fig:fdim_sampled_CRs}
\end{figure}

In Fig.~\ref{fig:fdim_sampled_CRs}, we investigate how the contamination rate from the Pang catalog changes the fractal dimension in five example clusters, which have different morphologies (tidal-tail and halo-types). We observe a small decline of the mean fractal dimension (represented by blue crosses in Fig.~\ref{fig:fdim_sampled_CRs}) when the contamination rate increases from 1\% to 5\%. However, this might be induced by low-number statistics, as the member number significantly drops when the contamination rate is below 5\%. On the other hand, the fractal dimension increases significantly as the contamination rate surpasses $15\%$. This is mainly attributed to the inclusion of more field stars, which generates artificial halo-like substructures in the outskirts \citep{pang2022a}, resulting in an artificial increase of the fractal dimension. The Pang catalog adopted a 5\% contamination rate for membership. As can be seen in Fig.~\ref{fig:fdim_sampled_CRs}, the dispersion of the fractal dimension is the largest at 5\%, implying that clusters of different morphologies can be effectively distinguished via fractal dimension. Therefore, members with a 5\% contamination rate are appropriate for fractal dimension analysis.
 
%%%%%%%%%%%%%%%%%%%%%%%%%%%%%%%%%%%
%%%%%%%%%%%%%%%%%%%%%%%%%%%%%%%%%%%
%%%%%%%%%%%%%%%%%%%%%%%%%%%%%%%%%%%

\section{Fractal dimension of open clusters in the solar neighborhood}\label{sec:fdim_solar}

% introduce the purpose of the section and the 3 regions:
In this section, we analyze the results of the fractal dimension of open clusters in the solar neighborhood across different regions. We consider the entire cluster (the all-member region), the region inside the tidal radius ($r \leq r_{t}$), and the region outside the tidal radius ($r>r_{t}$). 

The tidal radii of the open clusters from both catalogs are computed  as
\begin{equation}\label{eq:rt}
    r_t=\left(\frac{G M_{\rm cl}}{2(A-B)^2}\right)^{\frac{1}{3}}
\end{equation}
\citep{pinfield1998}, where $G$ is the gravitational constant, $M_{\rm cl}$ is the total mass of the cluster (the sum of the masses of the individual cluster members), and parameters $A$ and $B$ are the Oort constants. Here we use $A = 15.3 \pm 0.4 $\,km\,s$^{-1}$\,kpc$^{-1}$ and $B = -11.9 \pm 0.4 $\,km\,s$^{-1}$\,kpc$^{-1}$ \citep{Bovy2017}. 

To avoid the bias of outliers, we set a threshold for the number of stars $n_{s}>30$ used to compute the fractal dimension. Regions with fewer than 30 stars are not assessed in our procedure.

\begin{figure}[tb!]
\centering
\resizebox{0.85\hsize}{!}{\includegraphics{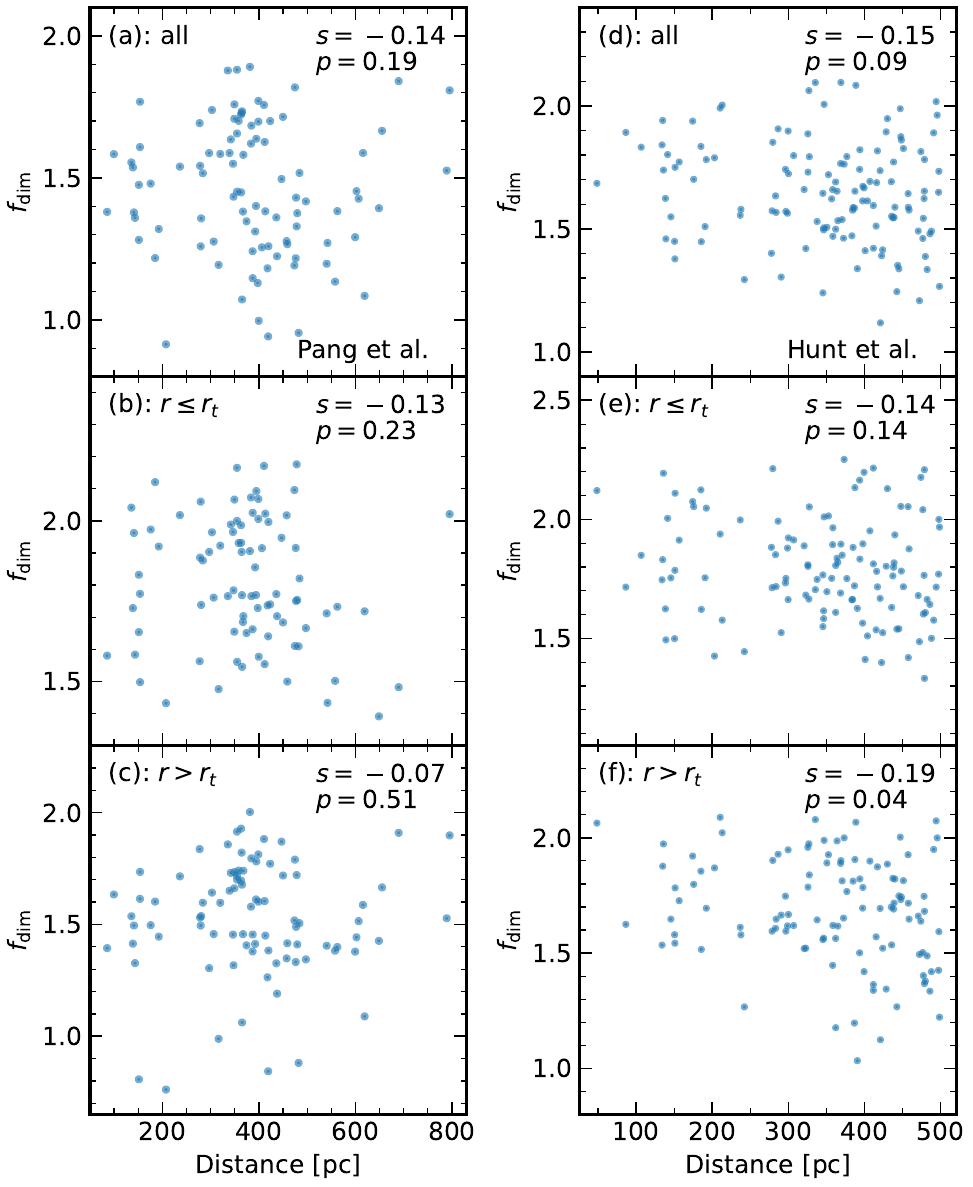}}
    \caption{The relation between fractal dimension and the corrected distance of the Pang catalog clusters (left panel) and the Hunt catalog clusters (right panel) in the solar neighborhood for the all-member region ((a) and (d)), $r \leq r_{t}$ ((b) and (e)), and $r > r_t$ ((c) and (f)). The quantity $s$ is Spearman’s rank correlation coefficient, and $p$ is the probability of the null hypothesis in the correlation test. A $p$-value below $0.05$ indicates a rejection of the null hypothesis. }
\label{fig:dist_fdim}
\end{figure}

Fig.~\ref{fig:dist_fdim} displays the fractal dimension of open clusters as a function of distance. We do not observe any correlation between the fractal dimension and distance among the Pang catalog clusters (panels (a) to (c)). Therefore, the distance correction we applied allows us to successfully recover the cluster morphologies, which would otherwise be distorted by the uncertainties of the parallax measurements.

A similar trend is observed in Hunt catalog clusters in the all-member region and $r \leq r_t$ regions (panels~(d) to (e)). However, a weak correlation is observed for Hunt clusters in $r > r_t$ region (panel~(f)), and the fractal dimension decreases as the distance increases. As a significant number of the Hunt clusters are elongated along the line of sight, the distance correction method is not optimal for correcting this morphology (see Figure~4 in \citet{pang2021a}). This indicates that a small bias will be introduced when we use the fractal dimension to quantify the extended regions of Hunt catalog clusters. We will address this effect in the analysis below.

%%%%%%%%%%%%%%%%%%%%%%%%%%%%%%%%%
%%%%%%%%%%%%%%%%%%%%%%%%%%%%%%%%%
%%%%%%%%%%%%%%%%%%%%%%%%%%%%%%%%%

\subsection{Distribution of fractal dimension in different regions} \label{sec:fdim_dist_solar}

% results for the three regions:

\begin{figure}[tb!]
\centering
\resizebox{0.95\hsize}{!}{\includegraphics{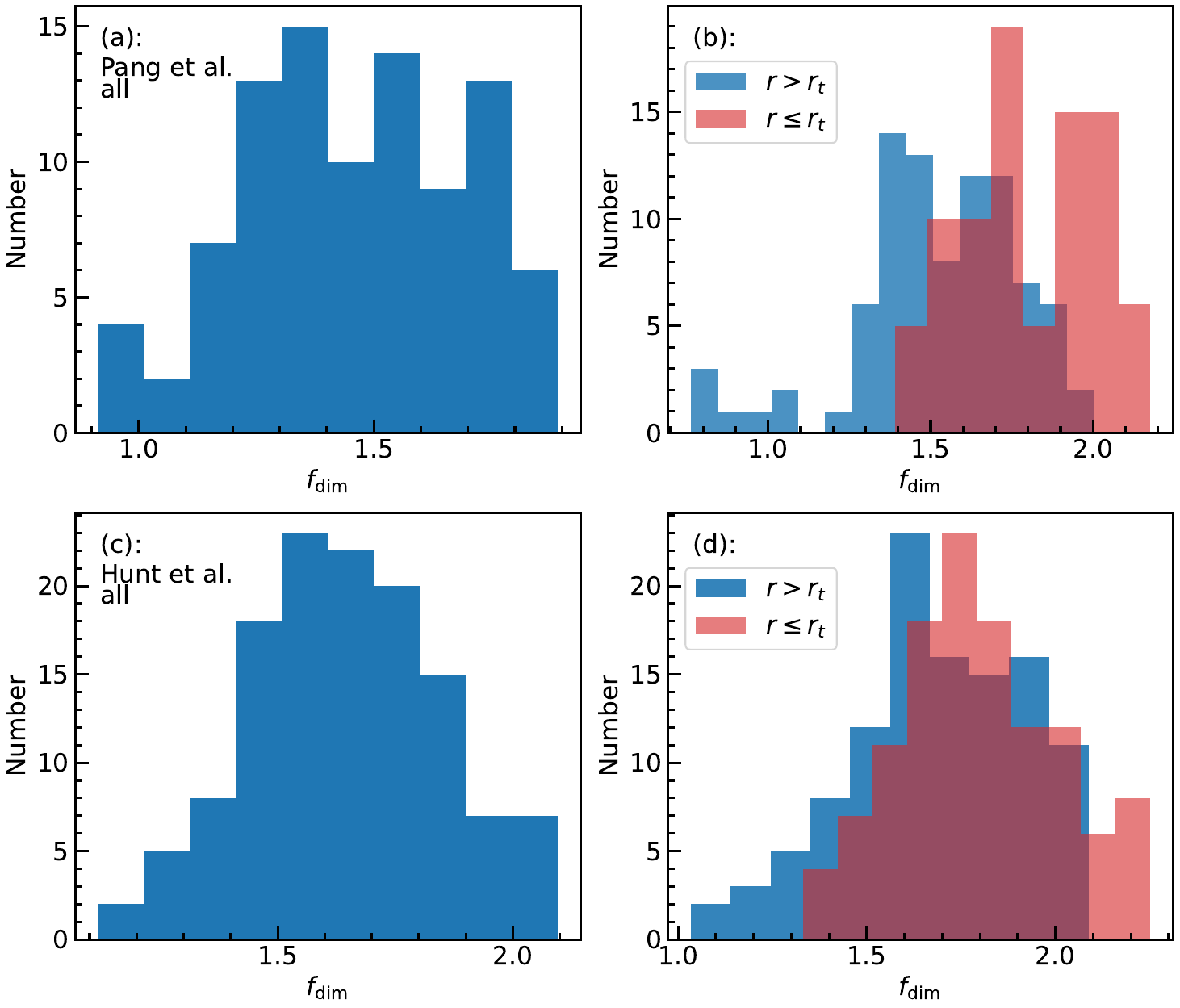}}
    \caption{(a): Histogram of fractal dimension for the Pang catalog clusters. (b): Overlapping histograms of fractal dimension for the Pang catalog clusters at $r \leq r_{t}$ (red) and $r > r_{t}$ (blue). (c): Histogram of fractal dimension for the Hunt catalog clusters. (d): Overlapping histograms of fractal dimension for the Hunt catalog clusters at $r \leq r_{t}$ (red) and $r > r_{t}$ (blue).}
\label{fig:fdim_hist_comb}
\end{figure}

The distribution of the fractal dimension of clusters from \citet{Pang_2024} (for the all-member region) is presented in Fig.~\ref{fig:fdim_hist_comb}~(a). The mean fractal dimension is $f_{\rm dim} = 1.46$. The results for the regions inside and outside the tidal radius are represented with the red and blue histograms in Fig.~\ref{fig:fdim_hist_comb}~(b). The mean fractal dimension for $r > r_{t}$ is around $1.53$, with a prominent peak around $f_{\rm dim} \approx 1.4$ and a secondary peak at $f_{\rm dim} \approx 1.7$. The mean fractal dimension for $r \leq r_{t}$ increases to $f_{\rm dim} = 1.81$, with two prominent peaks at $1.7$ and $2.0$.

In Fig.~\ref{fig:fdim_hist_comb}~(c) and (d), we display the distribution of the fractal dimension for three regions of clusters in the Hunt catalog clusters. The mean fractal dimension is approximately $1.65$ for the all-member region while that for $r > r_t$ and $r \leq r_t$ is $1.68$ and $1.80$ respectively. The distribution of fractal dimension is uni-modal for these three regions, while the all-member region is more symmetric. Both catalogs indicate that the fractal dimension for stars $r > r_{t}$ is smaller than that for $r \leq r_{t}$ (Fig.~\ref{fig:fdim_hist_comb}~(b) and (d)). However, the fractal dimension of Pang catalog clusters at $r > r_{t}$ covers a larger range, from roughly $0.8$ to $2.0$, compared to $1.0$ to $2.1$ in Hunt catalog clusters.
Unlike the Pang catalog clusters (Fig.~\ref{fig:fdim_hist_comb}~(b)), we find no significant difference between the fractal dimension distribution for $r \leq r_t$ (red histogram in Fig.~\ref{fig:fdim_hist_comb}~(d)) and $r > r_t$ (blue histogram in Fig.~\ref{fig:fdim_hist_comb}~(d)) of the Hunt catalog clusters except the slight shift of their peaks.

Among the three regions, the fractal dimension patterns are the same for both catalog clusters. The mean value at $r \leq r_{t}$ is the highest, whereas the mean for $r > r_{t}$ is lower, and the mean for the all-member region is the lowest.
This indicates that the region inside the tidal radius is more uniform, while that outside the tidal radius exhibits more substructure. This is in agreement with the conclusion from \cite{pang2022a}, which classifies the substructures outside the tidal radius into the filamentary, fractal, halo, and tidal tails. The lowest mean value for the all-member region is induced by the fixed fitting range of fractal dimension (Fig.~\ref{fig:fdim_examples}).

%%%%%%%%%%%%%%%%%%%%%%%%%%%%%%%%%
%%%%%%%%%%%%%%%%%%%%%%%%%%%%%%%%%
%%%%%%%%%%%%%%%%%%%%%%%%%%%%%%%%%

%%%%%%%%%%%%%%%%%%%%%%%%%%%%%%%%%
%%%%%%%%%%%%%%%%%%%%%%%%%%%%%%%%%
%%%%%%%%%%%%%%%%%%%%%%%%%%%%%%%%%

\subsection{Fractal dimension of morphological types}
\label{sec:fdim_type_solar}

\citet{pang2022a} visually categorized the morphology outside the tidal radius for 53 open clusters that show significant substructure (Col.~3 in Table~\ref{tab:fdim_Pang}) into four types: f1: filamentary, f2: fractal, h: halo, and t: tidal-tail. To be explicit, f1-type clusters are younger than $100$~Myr with unidirectional elongated substructures, while f2-type clusters are also younger than $100$~Myr but have fractal substructures. For clusters older than $100$~Myr, h-type clusters have a compact core with a low-density halo in the outskirts, whereas t-type clusters correspond to those with two tidal tails.

\begin{figure}[tb!]
\centering
\resizebox{0.85\hsize}{!}{\includegraphics{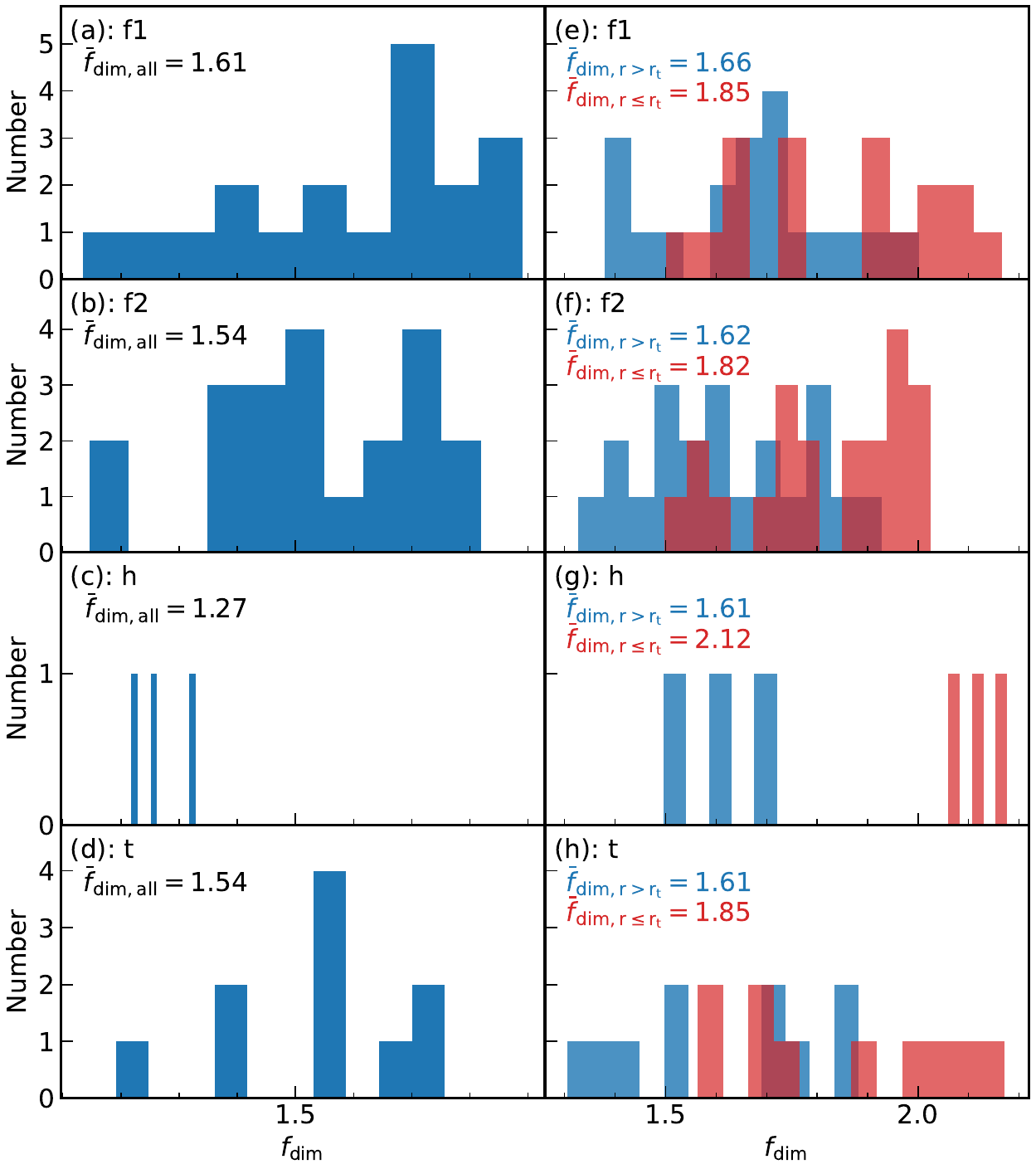}}
    \caption{Left: Histogram of fractal dimension for the all-member region in the 53 clusters from the Pang catalog, categorized by morphological type: (a): f1, (b): f2, (c): h, and (d): t. $\bar{f}_{\rm dim, all}$ represents the mean fractal dimension for each corresponding type. Right: Overlapping histograms of fractal dimension for the 53 clusters from the Pang catalog at $r \leq r_{t}$ (red) and at $r > r_{t}$ (blue), categorized by morphological type: (e): f1, (f): f2, (g): h, and (h): t. $\bar{f}_{\rm dim, r \leq r_t}$ and $\bar{f}_{\rm dim, r > r_t}$ represent the mean fractal dimension for each corresponding type of $r \leq r_{t}$ (red) and $r > r_{t}$ (blue). }
\label{fig:fdim_hist_type_93}
\end{figure}

Fig.~\ref{fig:fdim_hist_type_93} presents the histogram of the fractal dimensions for the morphology-classified clusters from \citet{pang2022a}. The fractal dimension of f1 and f2 types clusters (all-member region, panels (a) and (b)) exhibits the largest scatter, as reflected by its greater standard deviation of $0.213$ and $0.179$. In comparison, the standard deviation for halo-type and tidal-tail-type clusters have smaller values of $0.046$ and $0.167$, respectively. 
The distributions of fractal dimension for the f2-type and t-type clusters (all-member region, panels (b) and (d)) are very similar. The Kolmogorov-Smirnov test on these two distributions results in a p-value of $0.80$ (significantly larger than the threshold value of $0.05$), which means that these two distributions are very similar. From the right panels in Fig.~\ref{fig:fdim_hist_type_93}, we find that the mean fractal dimension values for the f1-type in $r \leq r_t$ and $r > r_{t}$ are $1.85$ and $1.66$, respectively; for the f2-type, they are $1.82$ and $1.62$; for the h-type, $2.12$ and $1.61$; and for the t-type, $1.85$ and $1.61$. From a comparison between the fractal dimensions of these four morphological types in the inner and outer regions, it is evident that the fractal dimension for $r \leq r_{t}$ is consistently greater than that for $r > r_{t}$, indicating a more uniform distribution within the tidal radius. The 53 morphologically classified clusters follow the same trend that was observed in all 93 clusters from the Pang catalog (Fig.~\ref{fig:fdim_hist_comb} (b)). The difference between $r > r_{t}$ and $r \leq r_{t}$ is most significant in halo- and tidal-tail-type clusters. These older clusters (h-type and t-type) have more unbound stars escaping in the outer halo-structure or tidal tails, respectively.

%%%%%%%%%%%%%%%%%%%%%%%%%%%%%%%%%
%%%%%%%%%%%%%%%%%%%%%%%%%%%%%%%%%
%%%%%%%%%%%%%%%%%%%%%%%%%%%%%%%%%

\subsection{Classification based on fractal dimension} \label{sec:fdim_classify}

In this section, we aim to quantitatively classify the morphology of open clusters via fractal dimension. The classification is based on the fractal dimension of three distinct regions as features: the all-member region ($f_{\rm dim, all}$), the region outside the tidal radius $r > r_{t}$ ($f_{\rm dim, r > r_t}$), and the region inside the tidal radius $r \leq r_{t}$ ($f_{\rm dim, r \leq r_t}$). 

We adopt the Gaussian Mixture Model (GMM) for the classifying process. GMM is characterized by its soft classifying approach, allowing each cluster to be associated with $k$ groups (we set $k \in [2,3,4,5]$ in GMM), and calculate the silhouette score corresponding to the $k$ groups for individual clusters. The $k$-group with the highest mean silhouette score in the cluster sample is selected to proceed for further analysis. Finally, we apply the Bayesian Information Criterion (BIC) to select the optimal number of groups, $k$, while avoiding over-fitting. The model with a low BIC value is considered as the optimal result. This method is flexible and can accommodate groups of diverse sizes. 

We present the average silhouette score in Table~\ref{tab:classify_results}. The silhouette scores quantify the similarity of a data point to its assigned group compared to other groups, with values in the range of $-1$ to $1$. An average silhouette score above $0.5$ suggests strong alignment with the assigned group and therefore good classification. When a data point's silhouette score approaches $-1$, the classification completely fails. 

\begin{table}[tb]
\centering
\caption{Classification with different feature combinations of star clusters in the solar neighborhood.}
\label{tab:classify_results}
\begin{tabular}{l@{\hspace{6pt}}l}
\hline \hline
Feature & $f_{\rm dim, all}$ \\
\hline
Groups & 2 \\
Silhouette Score & 0.5997 \\
Mean Age (Myr) & 83 / 259 \\
Clusters & 42 / 51 \\
\hline
Feature & $f_{\rm dim,r \leq r_t}$ \\
\hline
Groups & 2 \\
Silhouette Score & 0.6386 \\
Mean Age (Myr) & 115 / 244 \\
Clusters & 37 / 48 \\
\hline
\end{tabular}
\end{table}

The classification results for clusters from the Pang catalog are presented in Table~\ref{tab:classify_results}.
The fractal dimension for the all-member region and $r \leq r_t$ achieves a similar mean silhouette score $>0.5$ and classifies the Pang catalog clusters into two groups. There is an age difference between these two groups. The number of clusters in each group (see Table~\ref{tab:classify_results}) allows for a statistically significant comparison. The classification based on the fractal dimension for the $r > r_t$ region is ineffective. We are unable to divide clusters into different groups. Thus, the fractal dimension for $r > r_t$ alone does not appear to be a reliable feature for morphology classification.

As shown in \citet{pang2022a}, there are two kinds of clusters which both exhibit elongated substructures, but which have very different ages. The filamentary-type clusters are young ($<100$~Myr) and have inherited their filamentary structures from the filaments of molecular clouds. On the other hand, older clusters ($>100$~Myr), develop extended tidal tails and halos over time, due to the internal dynamical evolution and external Galactic tidal field. Therefore, the fractal dimension alone cannot be used to distinguish between different cluster shapes outside $r_t$ among fractal, filamentary, halo, and tidal-tail substructures \citep{pang2022a}.

%%%%%%%%%%%%%%%%%%%%%%%%%%%%%%%%%
%%%%%%%%%%%%%%%%%%%%%%%%%%%%%%%%%
%%%%%%%%%%%%%%%%%%%%%%%%%%%%%%%%%

\section{Discussion} \label{sec:fdim_discuss}

\subsection{Fractal dimension and open cluster dynamical evolution} \label{sec:fdim_dynamical}

\begin{figure}[tb!]
\centering
\resizebox{0.85\hsize}{!}{\includegraphics{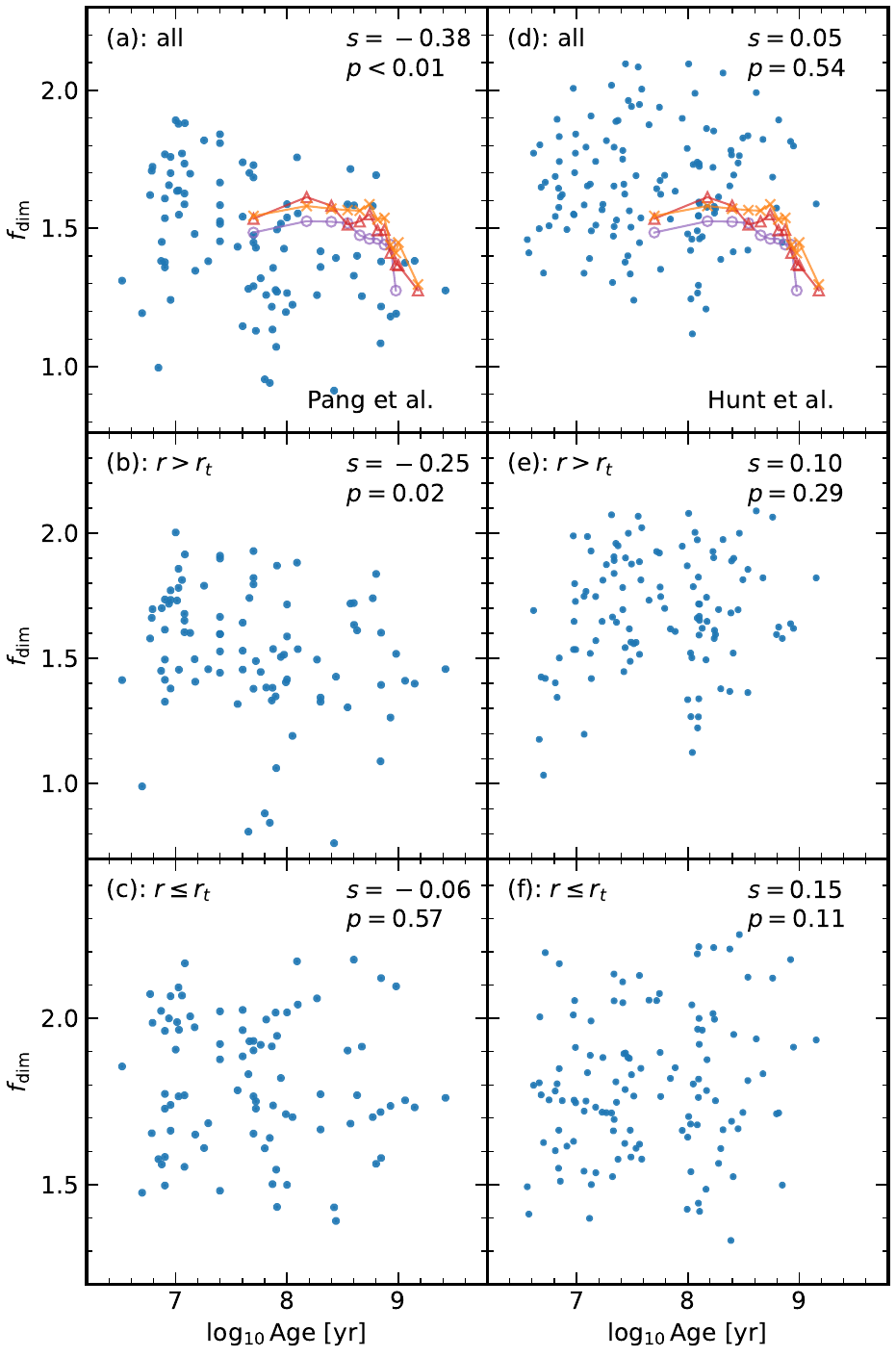}}
    \caption{The relation between the fractal dimension and the cluster age in  Pang catalog clusters (left panel) and in Hunt catalog clusters (right panel) in the solar neighborhood for the all-member region ((a) and (d)), $r > r_t$ ((b) and (e)), and $r \leq r_{t}$ ((c) and (f)). The colored open symbols and curves in panels (a) and (d) represent the evolution of the fractal dimension taken from the simulations by \citet{Akhmetali2024} with SFEs of $0.15$ (purple circles), $0.17$ (red triangles), and $0.20$ (orange crosses). The quantity $s$ is Spearman’s rank correlation coefficient, and $p$ is the probability of the null hypothesis in the correlation test. A $p$-value less than $0.05$ means rejection of the null hypothesis.
    }
\label{fig:logAge_fdim_ol}
\end{figure}

An earlier study by \citet{2009ApJ...696.2086S} first attempted to identify the relation between the fractal dimension of open clusters and cluster age, based on 2D positions. However, due to low-number statistics (8 clusters), they were unable to draw a definitive conclusion. 
In Fig.~\ref{fig:logAge_fdim_ol}~(a) to (c), we investigate the dependence of the fractal dimension on cluster age of the Pang catalog clusters. For the all-member region (panel (a)), the fractal dimension shows an anti-correlation with age, with the Spearman coefficient $s=-0.38$ and a $p$-value less than $0.01$. This implies that the morphology of the star clusters becomes more clumpy as the clusters evolve. This is expected from the dynamical evolution of open clusters, which results in the emergence of tidal tails and halos \citep{tarricq2022}, and therefore the fractal dimension decreases.

In Fig.\ref{fig:logAge_fdim_ol}~(a) and (d), we over-plot simulation results from \citet{Akhmetali2024}, who simulated star clusters with varying Star Formation Efficiencies (SFEs) of $0.15$ (purple circles), $0.17$ (red triangles), and $0.20$ (orange crosses), over ages ranging from $50$~Myr to $1500$~Myr. In their simulation, the fractal dimension evolves a little before 200~Myr. After 200~Myr, simulated star clusters start to develop tidal tails and lose members under the influence of the Galactic tidal field. Therefore, their fractal dimension declines significantly. The simulation results and our observations both cover a similar range of fractal dimensions and exhibit a generally negative trend.
The fractal dimension for Pang clusters at the region $r > r_{t}$ (Fig.~\ref{fig:logAge_fdim_ol}~(b)) also shows a declining trend with age ($s=-0.25$), similar to that of the all-member region.

On the other hand, for stars at $r \leq r_{t}$ in Pang clusters (panel (c) in Fig.~\ref{fig:logAge_fdim_ol}) the fractal dimension and the age show no dependence, with the Spearman coefficient $s$ being only $-0.06$, along with a $p$-value $0.57$. In the first 10\, Myr, a significant amount of crossing times have passed inside $r \leq r_t$, gravitational
interactions of member stars can effectively erase substructures. Therefore, further increases in the fractal dimension are asymptotically small. Given the scatter in the fractal dimension, we are unable to statistically detect the potentially slight relationship.

However, no correlation is observed between the fractal dimension and cluster age for the Hunt catalog clusters, as shown in Fig.~\ref{fig:logAge_fdim_ol}~(d) to (f). The bias we observed, as well as the weak dependence of fractal dimension on distance for Hunt clusters at $r>r_t$ (Fig.~\ref{fig:dist_fdim} (f)), might dilute the true dynamical evolution signature reflected by fractal dimension.

\subsection{Fractal dimension and star formation in open clusters} \label{sec:fdim_starformation}

\begin{figure}[tb!]
\centering
\resizebox{0.85\hsize}{!}{\includegraphics{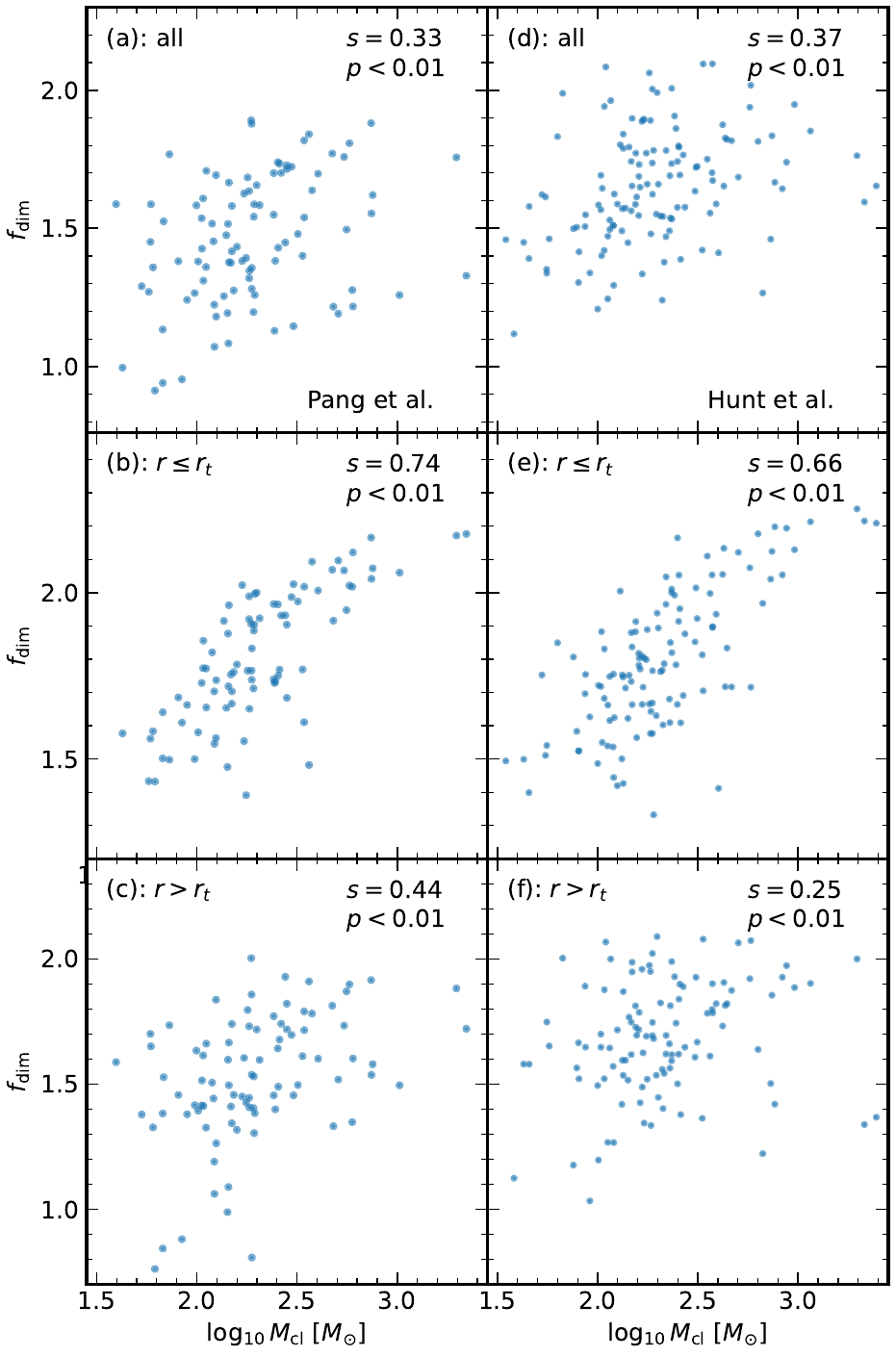}}
    \caption{The relation between fractal dimension and the observed total cluster mass in Pang catalog clusters (left panels) and Hunt catalog clusters (right panels) in the solar neighborhood for the all-member regions ((a) and (d)), for $r \leq r_{t}$ ((b) and (e)), and for $r > r_t$ ((c) and (f)). The quantity $s$ is Spearman’s rank correlation coefficient, and $p$ is the probability of the null hypothesis in the correlation test. A $p$-value less than $0.05$ means rejection of the null hypothesis.}
\label{fig:logMass_fdim_ol}
\end{figure}

Star formation in the solar neighborhood has been found to be hierarchical \citep{pang2022a}. We attempt to constrain the outcome of the star formation process in the solar neighborhood using the fractal dimension.  Fig.~\ref{fig:logMass_fdim_ol} illustrates the relationship between fractal dimension and cluster observed total mass, $M_{\rm cl}$, which is obtained by summing up the individual stellar masses provided by both catalogs. A pronounced correlation is observed between fractal dimension and cluster mass: $f_{\rm dim}$ for all three regions in the clusters of both catalogs increases with cluster mass. This trend is most prominent at $r \leq r_t$. 

These results align with the framework of hierarchical star formation. Through the conveyor belt mechanism, stars along filaments are transferred into dense hub regions via infalling flows, and these infalling stellar groups merge \citep{vazquez-semadeni_hierarchical_2016, trevino-morales2019}. 
The stellar groups formed in the filaments have a smaller SFE, and therefore cannot remain bound after gas expulsion. These low-mass groups inherited their filamentary shape from filaments and are characterized by lower fractal dimensions.  
While they merge to form more massive, and more centrally concentrated structures, the fractal dimension increases \citep{Schmeja2006}.  This correlation is particularly strong within $r \leq r_t$, where mergers are taking place. The higher value of $f_{\rm dim}$ among more massive clusters supports the hypothesis that they are probably formed hierarchically.

\subsection{Fractal dimension and galactic structures} \label{sec:fdim_galac}

\begin{figure*}[tb!]
\centering
\resizebox{0.8\hsize}{!}{\includegraphics{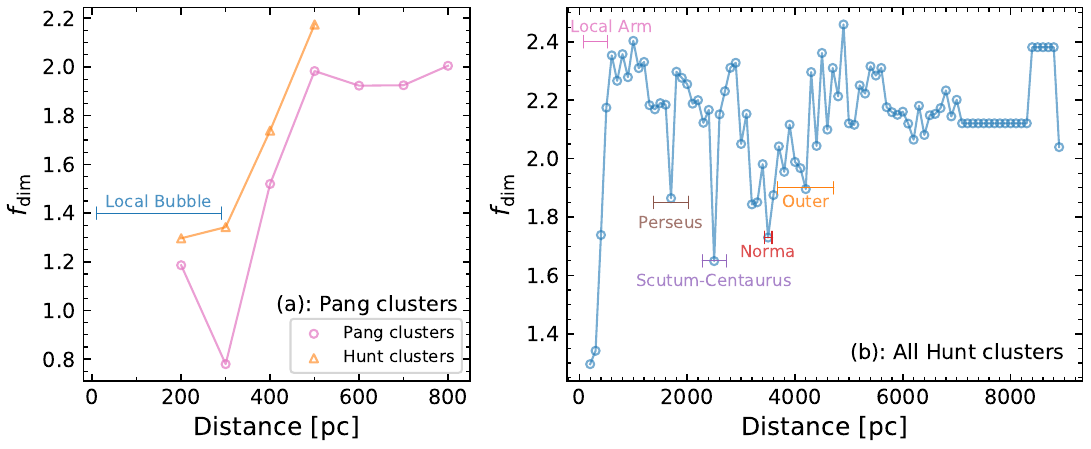}}
    \caption{
    Fractal dimension for the spatial distribution of open clusters vs. distance (a): in the solar neighborhood and (b): in the Galaxy. Note that the fractal dimension here is calculated based on the spatial distribution of individual clusters, rather than the internal distribution of member stars within each cluster, as discussed earlier. A cluster is represented as a single point in the 3D space, which is the median position of all stellar members. (a): Pink circles and orange triangles represent the fractal dimensions for the distribution of the Pang catalog clusters and Hunt catalog clusters, respectively. The blue bar indicates the extent of the Local Bubble. (b): Blue circles show the fractal dimension for the distribution of 3457 Hunt catalog open clusters across the Galaxy. The approximate locations and ranges of the Galactic arms are indicated in pink (Local Arm), brown (Perseus Arm), 
    purple (Scutum-Centaurus Arm), red (Norma Arm), and orange (Outer Arm). }
\label{fig:fdim_cluster_dist}
\end{figure*}

The analysis of the transition in the fractal dimension has been proposed by \citet{sanchez2010} to determine the typical size of star complexes in galaxies. On smaller scales in a galaxy, for example, star-forming regions, the fractal dimension is lower for the spatial distribution of star complexes (groups of star clusters and stellar groups), which resemble the fractal star formation cores induced by turbulence in giant molecular clouds. When we include stellar objects (from stars to star clusters) within a larger scale, the spatial distribution is consistent with a nearly uniform distribution. For example, the Galactic interstellar medium (ISM) has a universal fractal dimension $2.3 \pm 0.3$ \citep{elmegreen1996}, which represents a uniform distribution.

Below, we quantify the clustering strength of star clusters in the Milky Way via fractal dimension and aim to search for the transition of fractal dimension, which is induced by Galactic structures where stars are being formed. 
We adopt the median position of all members as the location of each cluster and calculate the fractal dimension for the distribution of open clusters within different distances. 
Fig.~\ref{fig:fdim_cluster_dist}~(a) displays the fractal dimension for the spatial distribution of star clusters from both Pang (pink circles) and Hunt (orange triangles) catalogs in the solar neighborhood. 
The computation of fractal dimension is only carried out for distances beyond 200\,pc since there are fewer than 10 clusters within 100\,pc. The computation is carried out with an interval of 100\,pc. At a distance of $\sim 300$\,pc, we observe a transition of $f_{\rm dim}$ in both catalogs. This distance is close to the edge of the Local Bubble \citep{O’Neill_2024}, which is a cavity characterized by high-temperature, low-density plasma, surrounded by a shell of dust and gas \citep{Donald1987}. Almost all star formation in the solar vicinity occurs on the surface of the Local Bubble \citep{Zucker2022}. Beyond 300 pc, the fractal dimension increases dramatically. The plateau observed in the fractal dimension above 500\,pc for the Pang catalog clusters is due to a very small number of clusters beyond 500\,pc.

In our Galaxy, the star-forming regions are concentrated along the spiral arms. To probe the larger-scale star formation structures in the Galaxy, we make use of all the open clusters in the Hunt catalog (a total of 3457 open clusters). 
Fig.~\ref{fig:fdim_cluster_dist}~(b) presents the fractal dimension for the distribution of these clusters across the Galaxy, computed with a 100\,pc interval, consistent with the calculation inside the solar neighborhood. At the distance of 600\,pc, the fractal dimension of open clusters approaches the value of Galactic ISM $f_{\rm dim}\sim2.3$ \citep{elmegreen1996}. This is the location at the edge of the Local Arm \citep{reid2019}, beyond which star formation and the spatial density of open clusters decline. 
We also notice a rapid drop in the fractal dimension from $2.3$  to $1.7-1.9$ at some locations, reaching several local minima. This occurs when the distribution of clusters approaches spiral arms, where the concentration of open clusters is higher. A comparison with the location of Galactic spiral arms \citep{reid2019, Hao2021} indicates that the local minima of $f_{\rm dim}$ correspond to the locations of the Perseus Arm (at $\sim$1700 pc), the Scutum-Centaurus Arm ($\sim$2500 pc), the Norma Arm ($\sim$3500 pc), and the Outer Arm ($\sim$ 4200\,pc). Therefore, the transition of the fractal dimension for the spatial distribution of open clusters provides a unique perspective to trace Galactic structures. It successfully recovers the small-scale structure inside the Solar neighborhood (the Local Bubble and the Local Arm) to larger structures in the Galaxy (the different spiral arms).

\section{Summary}\label{sec:sum}

We determine the fractal dimension for open clusters in the solar neighborhood using 3D position from Gaia DR\,3. Our cluster sample includes 93 clusters from the catalog of \citet{Pang_2024}, and 127 clusters from the catalog of \citet{Hunt2024}, mainly within 500~pc from the Sun. The box-counting method \citep{GRASSBERGER1983224} is applied to compute the fractal dimension of these clusters for three distinct regions: the all-member region; the region $r \leq r_t$ and the region $r > r_t$, where $r_t$ is the tidal radius. The open clusters can be effectively classified by their fractal dimension using the GMM algorithm. Based on the fractal dimension analysis of these open clusters, our results can be summarized as follows:

\begin{itemize}

\item The fractal dimension for the Pang catalog clusters in three distinct regions share a common peak around $f_{\rm dim}=1.7$. The fractal dimension is lower in the $r > r_t$ region, while it is highest within $r \leq r_t$.

\item We analyze the fractal dimensions of open clusters across the four morphological cluster types from \citep{pang2022a}: filamentary (f1), fractal (f2), halo (h), and tidal-tail (t), and conclude:

\begin{itemize}
    \item For the fractal dimension of the all-member region, f1-type clusters exhibit the largest variability and the highest mean value, while h-type clusters show the opposite. 

    \item Comparing the fractal dimensions of these morphological types across different regions, we find that the morphology inside the tidal radius is more uniform, while the overall morphology is more clumpy.
\end{itemize}

\item For the Hunt catalog clusters, the fractal dimension in all different regions exhibits a uni-modal distribution. The mean value of the fractal dimension, $f_{\rm dim} = 1.65$ is larger than that of the Pang catalog $f_{\rm dim} = 1.46$.

\item The GMM is utilized to classify clusters based on either the fractal dimension of all-member or $r \leq r_t$ regions, which divides the Pang catalog clusters into two groups with observable age differences. Fractal dimension for stars $r > r_t$  alone cannot properly classify cluster morphology.

\item We investigate the dynamical evolution of open clusters, star formation in open clusters, and Galactic structures, using the fractal dimension:

\begin{itemize}
    \item For the Pang clusters, the fractal dimension tends to decrease with age in both the all-member region and the $r > r_t$ region, which is consistent with the general evolutionary trend of the simulated clusters with varying SFEs described in \citet{Akhmetali2024}. In contrast, no correlation is observed for Hunt clusters across all three regions. 

    \item A strong correlation between the fractal dimension and cluster mass is observed in both catalogs across all three regions, with the relationship being more pronounced within $r \leq r_{t}$. This trend is closely linked to the hierarchical star formation process, in which massive clusters are built from the mergers of small filamentary groups. 

    \item The fractal dimension for the distribution of open clusters in both catalogs within the solar neighborhood increases rapidly beyond the shell of the Local Bubble (approximately 300\,pc). In the distribution of open clusters across the Galaxy (Hunt catalog, a total of 3457), local minima where $f_{\rm dim} < 1.9$ likely indicate the location of different spiral arms: the Local Arm ($\sim$600\,pc), the Perseus Arm ($\sim$1700\,pc), the Scutum-Centaurus Arm ($\sim$2500\,pc), the Norma Arm ($\sim$3500\,pc), and the Outer Arm ($\sim$4200\,pc).
    
\end{itemize}

\end{itemize}

Our study provides a novel approach to quantitatively analyze the 3D morphology of open clusters in the solar neighborhood through the fractal dimension. Further research will be conducted using more precise 3D spatial data that will be released from Gaia DR\,4. As a consequence of the uncertainty in the proper motion (PM) of Gaia DR\,3, PM measurements are strongly affected by unresolved binary stars \citep{pang2023}. Therefore, Gaia PMs cannot adequately probe the internal kinematics of star clusters, unless the interpretation of the data is corrected for the presence of unresolved binaries. The larger kinematic dataset will allow further exploration of the intricate relationship between morphology and cluster dynamics.

%TC:ignore

\begin{acknowledgements}
We thank the anonymous referee for providing helpful comments and suggestions that helped to improve this paper.
Xiaoying Pang acknowledges the financial support of the National Natural Science Foundation of China through grants 12173029 and 12233013. Antonella Vallenari acknowledges the support of INAF project CRA 1.05.23.05.19.
\\
This work made use of data from the European Space Agency (ESA) mission {\it Gaia} 
(\url{https://www.cosmos.esa.int/gaia}), processed by the {\it Gaia} Data Processing 
and Analysis Consortium (DPAC, \url{https://www.cosmos.esa.int/web/gaia/dpac/consortium}).
\end{acknowledgements}

%This study also made use of the SIMBAD database and the VizieR catalogue access tool, both operated at CDS, Strasbourg, France.

%------------------

%-------------------

%%%%%%%%%%%%%%%%%%%%%%%%%%%%%%%%%%%%%%%%%%%%%%%%%%%%%%%
%\newpage
\bibliographystyle{aa}
\bibliography{main}
%%%%%%%%%%%%%%%%%%%%%%%%%%%%%%%%%%%%%%%%%%%%%%%%%%%%%%%

\clearpage

\onecolumn

\begin{appendix}
\section{Tables}

% Switch to one-column mode for the longtable
%\onecolumn

\LTcapwidth=\textwidth

\begin{longtable}[1]{l c c c c c}
\caption{Fractal dimension for open clusters from \citet{Pang_2024}.\label{tab:fdim_Pang}} \\
\hline\hline
Cluster & Age (Myr) & Type & $f_{\rm dim}$ (all) & $f_{\rm dim}$ ($r \leq r_t$) & $f_{\rm dim}$ ($r > r_t$) \\
\hline
\endfirsthead

\caption{continued.} \\
\hline\hline
Cluster & Age (Myr) & Type & $f_{\rm dim}$ (all) & $f_{\rm dim}$ ($r \leq r_t$) & $f_{\rm dim}$ ($r > r_t$) \\
\hline
\endhead

\hline
\endfoot

\hline
\endlastfoot

$\mathrm{Alessi\ 20}$ & 9 & f1 & 1.63 & 1.55 & 1.60 \\
$\mathrm{Alessi\ 20\ gp1}$ & 12 & f2 & 1.27 & 1.50 & 1.42 \\
$\mathrm{Alessi\ 20\ isl1}$ & 100 &  & 1.70 & 1.74 & 1.77 \\
$\mathrm{Alessi\ 24}$ & 88 &  & 1.52 & 1.82 & 1.51 \\
$\mathrm{Alessi\ 3}$ & 631 & t & 1.69 & 1.56 & 1.84 \\
$\mathrm{Alessi\ 5}$ & 52 &  & 1.13 & 1.73 & - \\
$\mathrm{Alessi\ 62}$ & 691 &  & 1.08 & 1.72 & 1.09 \\
$\mathrm{Alessi\ 9}$ & 265 &  & 0.91 & 1.43 & 0.76 \\
$\mathrm{ASCC\ 105}$ & 74 & f1 & 1.13 & 1.50 & 1.38 \\
$\mathrm{ASCC\ 127}$ & 15 & f1 & 1.35 & 1.65 & 1.41 \\
$\mathrm{ASCC\ 16}$ & 10 &  & 1.55 & 1.97 & 1.45 \\
$\mathrm{ASCC\ 19}$ & 8 & f2 & 1.66 & 2.00 & 1.72 \\
$\mathrm{ASCC\ 32}$ & 25 &  & 1.81 & 2.02 & 1.90 \\
$\mathrm{ASCC\ 58}$ & 52 & f2 & 1.43 & 1.75 & 1.49 \\
$\mathrm{BH\ 164}$ & 65 &  & 1.26 & 2.00 & 1.38 \\
$\mathrm{BH\ 99}$ & 81 & f2 & 1.50 & 1.95 & 1.87 \\
$\mathrm{Blanco\ 1}$ & 100 & t & 1.54 & 2.02 & 1.72 \\
$\mathrm{Collinder\ 132\ gp1}$ & 25 &  & 1.67 & - & 1.67 \\
$\mathrm{Collinder\ 132\ gp2}$ & 25 &  & 1.84 & 1.48 & 1.91 \\
$\mathrm{Collinder\ 132\ gp3}$ & 25 &  & 1.45 & - & 1.44 \\
$\mathrm{Collinder\ 132\ gp4}$ & 25 &  & 1.53 & - & 1.53 \\
$\mathrm{Collinder\ 132\ gp5}$ & 50 &  & 1.29 & - & 1.38 \\
$\mathrm{Collinder\ 132\ gp6}$ & 100 &  & 1.59 & - & 1.59 \\
$\mathrm{Collinder\ 135}$ & 40 & f2 & 1.74 & 1.96 & 1.64 \\
$\mathrm{Collinder\ 140}$ & 50 & f2 & 1.68 & 1.77 & 1.80 \\
$\mathrm{Collinder\ 350}$ & 589 & t & 1.58 & 1.70 & 1.74 \\
$\mathrm{Collinder\ 69}$ & 13 & f1 & 1.70 & 2.01 & 1.60 \\
$\mathrm{Coma\ Berenices}$ & 700 & t & 1.38 & 1.58 & 1.39 \\
$\mathrm{Group\ X}$ & 400 & d & 1.58 & - & 1.63 \\
$\mathrm{Gulliver\ 21}$ & 275 &  & 1.39 & 1.39 & 1.43 \\
$\mathrm{Gulliver\ 6}$ & 7 & f1 & 1.38 & 2.02 & 1.45 \\
$\mathrm{Huluwa\ 1}$ & 12 & f1 & 1.88 & 2.17 & 1.92 \\
$\mathrm{Huluwa\ 2}$ & 11 & f1 & 1.77 & 2.07 & 1.81 \\
$\mathrm{Huluwa\ 3}$ & 10 &  & 1.64 & 2.09 & 1.78 \\
$\mathrm{Huluwa\ 4}$ & 10 & f1 & 1.64 & 1.99 & 1.73 \\
$\mathrm{Huluwa\ 5}$ & 7 & f1 & 1.45 & 1.56 & 1.70 \\
$\mathrm{IC\ 2391}$ & 50 &  & 1.48 & 1.65 & - \\
$\mathrm{IC\ 2602}$ & 45 &  & 1.28 & 1.83 & 0.81 \\
$\mathrm{IC\ 348}$ & 5 &  & 1.19 & 1.48 & 0.99 \\
$\mathrm{IC\ 4665}$ & 36 &  & 1.43 & 1.78 & 1.32 \\
$\mathrm{IC\ 4756}$ & 955 & t & 1.19 & 2.10 & 1.52 \\
$\mathrm{LP\ 2371}$ & 19 &  & 1.38 & 1.68 & 1.46 \\
$\mathrm{LP\ 2373}$ & 4 & f1 & 1.88 & 1.77 & 1.86 \\
$\mathrm{LP\ 2373\ gp1}$ & 10 & f1 & 1.76 & 2.07 & 1.73 \\
$\mathrm{LP\ 2373\ gp2}$ & 9 & f1 & 1.71 & 1.65 & 1.66 \\
$\mathrm{LP\ 2373\ gp3}$ & 6 & f1 & 1.72 & 1.99 & 1.70 \\
$\mathrm{LP\ 2373\ gp4}$ & 6 & f2 & 1.24 & 1.66 & 1.38 \\
$\mathrm{LP\ 2383}$ & 50 & f2 & 1.73 & 1.90 & 1.82 \\
$\mathrm{LP\ 2388}$ & 22 &  & 1.42 & 1.67 & 1.34 \\
$\mathrm{LP\ 2428}$ & 200 &  & 1.36 & 1.77 & 1.33 \\
$\mathrm{LP\ 2429}$ & 1150 & t & 1.38 & 1.75 & 1.41 \\
$\mathrm{LP\ 2439}$ & 25 & f2 & 1.52 & 1.88 & 1.60 \\
$\mathrm{LP\ 2441}$ & 75 & f2 & 1.36 & 1.74 & 1.54 \\
$\mathrm{LP\ 2442}$ & 14 & f2 & 1.54 & 1.73 & 1.41 \\
$\mathrm{LP\ 2442\ gp1}$ & 8 & f2 & 1.38 & 1.96 & 1.50 \\
$\mathrm{LP\ 2442\ gp2}$ & 8 & f2 & 1.36 & 1.58 & 1.33 \\
$\mathrm{LP\ 2442\ gp3}$ & 8 & f2 & 1.61 & 1.77 & 1.61 \\
$\mathrm{LP\ 2442\ gp4}$ & 8 & f2 & 1.77 & 1.50 & 1.74 \\
$\mathrm{LP\ 2442\ gp5}$ & 8 & f2 & 1.48 & 1.97 & 1.50 \\
$\mathrm{Mamajek\ 4}$ & 371 & t & 1.71 & 1.68 & 1.72 \\
$\mathrm{NGC\ 1901}$ & 850 &  & 1.18 & 1.74 & 1.26 \\
$\mathrm{NGC\ 1977}$ & 3 &  & 1.31 & 1.86 & 1.41 \\
$\mathrm{NGC\ 1980}$ & 5 &  & 1.62 & 2.07 & 1.58 \\
$\mathrm{NGC\ 2232}$ & 25 & f1 & 1.58 & 1.92 & 1.60 \\
$\mathrm{NGC\ 2422}$ & 73 &  & 1.22 & 1.92 & 1.33 \\
$\mathrm{NGC\ 2451A}$ & 58 &  & 1.32 & 1.92 & 1.45 \\
$\mathrm{NGC\ 2451B}$ & 50 & f2 & 1.45 & 1.93 & 1.93 \\
$\mathrm{NGC\ 2516}$ & 123 & t & 1.76 & 2.17 & 1.88 \\
$\mathrm{NGC\ 2547}$ & 40 & f2 & 1.15 & 2.03 & 1.46 \\
$\mathrm{NGC\ 3228}$ & 63 &  & 0.95 & 1.61 & 0.88 \\
$\mathrm{NGC\ 3532}$ & 398 & h & 1.33 & 2.18 & 1.72 \\
$\mathrm{NGC\ 6405}$ & 79 &  & 1.28 & 2.02 & 1.35 \\
$\mathrm{NGC\ 6475}$ & 186 & h & 1.26 & 2.06 & 1.50 \\
$\mathrm{NGC\ 6633}$ & 426 &  & 1.40 & 1.77 & 1.61 \\
$\mathrm{NGC\ 6774}$ & 2650 &  & 1.28 & 1.76 & 1.46 \\
$\mathrm{NGC\ 6991}$ & 1400 &  & 1.38 & 1.73 & 1.40 \\
$\mathrm{NGC\ 7058}$ & 80 &  & 1.07 & 1.55 & 1.06 \\
$\mathrm{NGC\ 7092}$ & 350 & t & 1.59 & 1.90 & 1.30 \\
$\mathrm{Pleiades}$ & 125 & t & 1.55 & 2.04 & 1.54 \\
$\mathrm{Praesepe}$ & 700 & h & 1.22 & 2.12 & 1.60 \\
$\mathrm{Roslund\ 5}$ & 97 & f2 & 1.20 & 1.71 & 1.40 \\
$\mathrm{RSG\ 7}$ & 70 &  & 0.94 & 1.64 & 0.84 \\
$\mathrm{RSG\ 8}$ & 18 & f2 & 1.82 & 1.61 & 1.79 \\
$\mathrm{Stephenson\ 1}$ & 46 & f1 & 1.70 & 1.93 & 1.74 \\
$\mathrm{Stock\ 1}$ & 470 &  & 1.22 & 1.70 & 1.19 \\
$\mathrm{Stock\ 12}$ & 112 &  & 1.25 & 1.91 & - \\
$\mathrm{Stock\ 23}$ & 94 & f1 & 1.43 & - & 1.51 \\
$\mathrm{UBC\ 19}$ & 7 &  & 1.00 & 1.58 & - \\
$\mathrm{UBC\ 31}$ & 12 & f1 & 1.59 & - & 1.65 \\
$\mathrm{UBC\ 31\ gp1}$ & 12 & f1 & 1.89 & 1.91 & 2.00 \\
$\mathrm{UBC\ 31\ gp2}$ & 10 & f1 & 1.73 & 1.77 & 1.68 \\
$\mathrm{UBC\ 7}$ & 40 & f2 & 1.54 & 1.89 & 1.53 \\
$\mathrm{UPK\ 82}$ & 81 &  & 1.27 & 1.43 & - \\

\end{longtable}

\begin{flushleft}
\tablefoot{The ages of the clusters are obtained from \citet{pang2021a, pang2021b, pang2022a, pang2022c, Pang_2024}, derived via the PAdova and TRieste Stellar Evolution Code (PARSEC) isochrone fitting. Col.~3 lists the morphological type from \citet{pang2022a}. Cols.~4-6 present the fractal dimensions of open clusters for the all-member region, for the region inside the tidal radius $r \leq r_t$, and for the region outside $r > r_t$, respectively. The fractal dimension of some open clusters in specific regions is considered invalid due to their low number of member stars (see Sect.~\ref{sec:fdim_solar}).}
\end{flushleft}

\FloatBarrier

\begin{table}
\centering
\caption{Fractal dimension for open clusters from \citet{Hunt2024}.\label{tab:fdim_Hunt_10}}
\begin{tabular}{l c c c c}
\hline\hline
Cluster & $\rm log_{10}(Age/Myr)$ & $f_{\rm dim}$ (all) & $f_{\rm dim}$ ($r \leq r_t$) & $f_{\rm dim}$ ($r > r_t$) \\
\hline

$\mathrm{Alessi\ 10}$ & 8.03 & 1.24 & 1.54 & 1.27 \\
$\mathrm{Alessi\ 20}$ & 6.83 & 1.89 & 1.80 & 1.34 \\
$\mathrm{Alessi\ 24}$ & 8.00 & 1.48 & 1.64 & 1.34 \\
$\mathrm{Alessi\ 34}$ & 7.31 & 2.02 & 1.72 & 2.07 \\
$\mathrm{Alessi\ 36}$ & 7.46 & 1.40 & 1.88 & - \\
$\mathrm{Alessi\ 3}$ & 8.80 & 1.57 & 1.71 & 1.60 \\
$\mathrm{Alessi\ 5}$ & 7.75 & 1.67 & 1.90 & 1.78 \\
$\mathrm{Alessi\ 96}$ & 8.32 & 2.06 & 1.67 & 1.97 \\
$\mathrm{Alessi\ 9}$ & 8.61 & 1.99 & 1.94 & 2.09 \\
$\mathrm{ASCC\ 101}$ & 8.28 & 1.61 & 1.56 & 1.70 \\

\hline
\end{tabular}
\begin{flushleft}
\tablefoot{A machine-readable version of this table is available online. Only the first ten clusters are presented as examples. The full table is available in its entirety in machine-readable form.}
\end{flushleft}
\end{table}

\end{appendix}

%TC:endignore

\end{document}